\documentstyle[12pt]{article}
\newcommand{\bd}{\begin{document}}
\newcommand{\ed}{\end{document}}
\newcommand{\bc}{\begin{center}}
\newcommand{\ec}{\end{center}}
\newcommand{\be}{\begin{eqnarray}}
\newcommand{\ee}{\end{eqnarray}}
\renewcommand{\thefootnote}{\alph{footnote}}
\newcommand{\se}{\section}
\newcommand{\sse}{\subsection}
\newcommand{\bi}{\bibitem}
\def\figcap{\section*{Figure Captions\markboth
     {FIGURECAPTIONS}{FIGURECAPTIONS}}\list
     {Figure \arabic{enumi}:\hfill}{\settowidth\labelwidth{Figure 999:}
     \leftmargin\labelwidth
     \advance\leftmargin\labelsep\usecounter{enumi}}}
\let\endfigcap\endlist \relax
\input psfig

\begin{document}

\begin{titlepage}

 \vskip 0.5in
 \null
\begin{center}
 \vspace{.15in}
{\Large {\bf What can we learn from $B\rightarrow$ $\rho
(\omega)\,K$ decays?
}}\\
\vspace{1.0cm}  \par
 \vskip 2.1em
 {\large
\begin{tabular}[t]{c}
{\bf Chuan-Hung Chen${}^{a,b}$}
\\
\\
\\
{\sl ${}^a$Department of Physics, National Cheng Kung University,
}
\\  {\sl  Tainan, Taiwan 701, Republic of China }
\\
\\
   {\sl ${}^b$Institute of Physics, Academia Sinica, Taipei,}
\\   {\sl  $\ $Taiwan 115, Republic of China }

\end{tabular}}
 \par \vskip 5.3em

 {\Large\bf Abstract}

\end{center}
We show that in the PQCD approach the $B\rightarrow \rho(\omega)
K$
 decays display not only the dynamical penguin enhancement,
 the mechanism responsible for the large branching ratios (BRs) of $B\rightarrow \phi K$ decays, but
 also the importance of annihilation contributions.
   We find that the CP asymmetries (CPAs) of $B\rightarrow
 \rho^{\pm}K^{\mp}$, $B^{\mp}\rightarrow \rho^0(\omega)K^{\mp}$
 (class I) are all over $50\%$.\\
 %The predicted
%$B^{\mp}\rightarrow \omega K^{\mp}$ BR of $3.22\times 10^{-6}$
% for $\gamma(\phi_{3})\simeq 72^0$  is
 %the same as the reported value of CLEO.\\

\noindent{PACS number(s): 12.38.BX, 12.38.QK, 13.25.HW }

\end{titlepage}

The study of charmless $B$ meson decays has an enormous progress, since many
modes, such as $B\rightarrow \pi \pi $ and $B\rightarrow \pi K$, were
measured by CLEO \cite{pipiCLEO}, as well as by BABAR \cite{pipiBABAR} and
BELLE \cite{pipiBELLE}. Via the search of $B$ decays, we can not only test
the origin of CP violation in standard model (SM), which is the consequence
of the Cabibbo-Kobayashi-Maskawa (CKM ) quark-mixing matrix \cite{CKM}, but
also verify various QCD approaches, proposed to deal with hadronic effects
in exclusive decays. Recently, the BRs of the modes involving a vector (V)
meson, such as $B\rightarrow \rho (\omega )\pi $ measured by CLEO \cite
{CLEO1} and $B^{\pm }\rightarrow \phi K^{\pm }$ by the $B$ factories,
\begin{eqnarray*}
B^{\pm }\rightarrow \phi K^{\pm } &=&(5.5_{-1.8}^{+2.1}\pm 0.6)\times
10^{-6}\quad ({\rm CLEO\ }\cite{CLEO2}), \\
&=&(7.7_{-1.4}^{+1.6}\pm 0.8)\times 10^{-6}\quad ({\rm BABAR\ }\cite{BABAR}),
\\
&=&(11.2_{-2.0}^{+2.2}\pm 1.4)\times 10^{-6}\ ({\rm BELLE\
}\cite{BELLE})
\end{eqnarray*}
have been also observed. Although the present results of $B^{\pm
}\rightarrow \phi K^{\pm }$ among the $B$ factories do not match well each
other, more accurate measurements will be obtained in the near future.
Nevertheless, the theoretical predictions are not consistent either; for
instance, the modified perturbative QCD (MPQCD) predicts $%
(10.2_{-2.1}^{+3.9})\times 10^{-6}$ \cite{Mishima,CKL}, but QCD
factorization (QCDF) \cite{BBNS} gives $(4.3_{-1.4}^{+3.0})\times
10^{-6}$ \cite{CY}. Therefore, an interesting question is raised:
besides $B\rightarrow \phi K$ decays, can we find other processes,
that distinguish which QCD approach is proper and correct? or what
processes can support either of them? Following the speculation of
$B\rightarrow VP$ decays, $P$ being a pseudoscalar meson, we
find that it is important to investigate $B\rightarrow V^{\prime }K$ decays (%
$V^{\prime }=\rho $ and $\omega $), because many properties, such as CPAs of
over $50\%$ and essential annihilation contributions, are remarkable.
Although the $V^{\prime }$ meson carries spin degrees of freedom, only the
longitudinal component contributes. Thus, in order to understand the
properties of $B\rightarrow V^{\prime }K$, it is useful to combine the
analysis of $B\rightarrow \pi ^{-}K^{+}$ and $B^{+}\rightarrow \phi K^{+}$
decays, all of which are described by the $b\to s$ transition. Note that the
pion can be regarded as the longitudinal component of a $\rho $ meson.

The effective Hamiltonian for decays with the $b\to s$ transition is given
by
\begin{equation}
H_{{\rm eff}}=\frac{G_{F}}{\sqrt{2}}\sum_{q^{\prime }=u,c}V_{q^{\prime
}}\left[ C_{1}(\mu ){\cal O}_{1}^{(q^{\prime })}+C_{2}(\mu ){\cal O}%
_{2}^{(q^{\prime })}+\sum_{i=3}^{10}C_{i}(\mu ){\cal O}_{i}\right]
\label{eff}
\end{equation}
where $V_{q^{\prime }}=V_{q^{\prime }s}^{*}V_{q^{\prime }b}$ are the
products of the CKM matrix elements, $C_{i}(\mu )$ are the Wilson
coefficients (WCs) and ${\cal O}_{i}$ correspond to the four-quark
operators. The explicit expressions of $C_{i}(\mu )$ and ${\cal O}_{i}$ can
be found in Ref.\cite{BBL}.
%%%The structures of ${\cal O}_{3,4,5,6}$, generated by the QCD penguin, are the same
%%%as those of ${\cal O}_{9,10,7,8}$ from the electroweak (EW)
%%%penguin, respectively. Except the different color flows between
%%%$O_{2i-1}$ and $O_{2i}$ $(i=1,2,\cdots ,5)$,
%%%%they have the same structure.
We can define the evolution variables as
\begin{eqnarray}
a_{1} &=&C_{1}+\frac{C_{2}}{N_{c}},\,a_{2}=C_{2}+\frac{C_{1}}{N_{c}}%
,\,a_{1}^{\prime }=\frac{C_{2}}{N_{c}},\,a_{2}^{\prime }=\frac{C_{1}}{N_{c}},
\nonumber \\
a_{3,4}^{(q)} &=&C_{3,4}+\frac{3e_{q}}{2}C_{9,10}+a_{3,4}^{\prime
(q)},\,a_{3,4}^{\prime (q)}=\frac{C_{4,3}}{N_{c}}+\frac{3e_{q}}{2N_{c}}%
C_{10,9},  \nonumber \\
a_{5,6}^{(q)} &=&C_{5,6}+\frac{3e_{q}}{2}C_{7,8}+a_{5,6}^{\prime
(q)},\,a_{5,6}^{\prime (q)}=\frac{C_{6,5}}{N_{c}}+\frac{3e_{q}}{2N_{c}}%
C_{8,7}  \label{ew}
\end{eqnarray}
where the superscripts $q$ represent the light u, d and s quarks.
%%%%All decay amplitudes related to the weak dynamical interactions depend on $%
%%%a_{j}^{(q)}$ and $a_{j}^{\prime (q)}$ $(j=1,2,\cdots ,6)$.
Note that $a_{2}$ is larger than $a_{1}$ and nonfactorizable effects are
only associated with the color suppressed variables $a_{j}^{\prime (q)}$.

For a simple analysis, we first concentrate on $B\rightarrow \rho ^{-}K^{+} $
decay, because it contains the common properties of $B\rightarrow V^{\prime
}K$ decays. Hence, based on Eq. (\ref{eff}), the decay amplitudes for $%
B\rightarrow \pi ^{-}K^{+}$, $B^{+}\rightarrow \phi K^{+}$ and $B\rightarrow
\rho ^{-}K^{+}$ are written as
\begin{eqnarray}
A\left( B\rightarrow \pi ^{-}K^{+}\right) &=&f_{K}\left[ V_{t}^{*}\left(
a_{4}^{(u)}+2r_{K}a_{6}^{(u)}\right) -V_{u}^{*}a_{2}\right]  \nonumber \\
&&\times F_{e}^{B\pi }+2f_{B}V_{t}^{*}a_{6}^{(d)}F_{a6}^{\pi K}+...,
\label{pik} \\
A\left( B^{+}\rightarrow \phi K^{+}\right) &=&f_{\phi }V_{t}^{*}\left(
a_{3}^{(s)}+a_{4}^{(s)}+a_{5}^{(s)}\right) F_{e}^{BK}  \nonumber \\
&&+2f_{B}V_{t}^{*}a_{6}^{(u)}F_{a6}^{\phi K}+...,  \label{phik} \\
A\left( B\rightarrow \rho ^{-}K^{+}\right) &=&f_{K}\left[ V_{t}^{*}\left(
a_{4}^{(u)}-2r_{K}a_{6}^{(u)}\right) -V_{u}^{*}a_{2}\right]  \nonumber \\
&&\times F_{e}^{B\rho }+2f_{B}V_{t}^{*}a_{6}^{(d)}F_{a6}^{\rho K}+...
\label{rhok}
\end{eqnarray}
with $r_{K}=m_{K}^{0}/M_{B}$, $m_{K}^{0}$ being the chiral symmetry breaking
(CSB) parameter associated with the kaon. We present only the dominant and
subdominant effects with nonfactorizable effects neglected. However, the
complete effects will be included in our numerical calculations. $f_{K}$, $%
f_{\phi }$ and $f_{B}$ are the decay constants of $K$, $\phi $ and $B$
mesons, respectively. $F_{e}^{BP(V)}$ denote the $B\rightarrow P(V)$ form
factors, and $F_{a6}^{P(V)P}$ are the time-like form factors induced by the
annihilation contributions from the $(S-P)(S+P)$ four-quark operators. As
for $F_{a4}^{P(V)P}$ generated by the $(V-A)(V-A)$ operators, due to
helicity suppression, we do not discuss them. We emphasis that the
expressions of $a_{i}^{(q)}F_{e[a]}^{P(V)P}$ in Eqs. (\ref{pik})-(\ref{rhok}%
) are given only for the short-hand convenience. In the MPQCD approach, they
are convolutions of $a_{i}^{(q)}$ and $F_{e[a]}^{P(V)P}$ actually. If we
apply the idea of QCDF to MPQCD, in the viewpoint of numerical estimation,
the $\mu $ scale of $a_{i}^{(q)}(\mu )$ could be fixed at around the scale
of $1.7$ GeV. The scale is around $2.7$ GeV in QCDF \cite{CKL}.

Although $C_{2}$ is one order of magnitude larger than $C_{4}$ and
$C_{6}$, which are the first two largest WCs of QCD penguin, the
penguin emission topologies in the $\pi ^{-}K^{+}$ and $\phi
K^{+}$ modes are the dominant effects due to the suppression of
CKM matrix element $V_{u}\approx 1.6A\lambda ^{5}e^{-i\gamma }$.
The tree and penguin annihilation are subdominant. If taking
$\gamma\simeq 90^0$, the numerical value of Eq. (\ref
{pik}) is larger than that of Eq. (\ref{phik}) by an approximate factor of $%
\sqrt{1.8}$ and $\sqrt{3.1}$ with $\mu $ being fixed at the different scales
in MPQCD and QCDF, respectively. As known, the world average BR of $%
B\rightarrow \pi ^{\mp }K^{\pm }$ is $(17.2\pm 1.5)\times 10^{-6}$ \cite
{pipiCLEO,pipiBABAR,pipiBELLE}. Following the above analysis, we expect the
BR of $\phi K^{\pm }$ mode is around $9.5\times 10^{-6}$ ($\mu =1.7$ GeV)
and $5.5\times 10^{-6}$ ($\mu =2.7$ GeV)$.$ Clearly, with the information of
$B\rightarrow \pi ^{\mp }K^{\pm }$ in the present experimental status, we
cannot determine the typical QCD scale for heavy-to-light decays, unless
more precision measurements on $B\rightarrow \phi K$ are given.

Different from $B\rightarrow \pi ^{-}K^{+}$ and $B^{+}\rightarrow
\phi K^{+}$ decays, in terms of Eq. (\ref{rhok}), we see that due
to the cancellation between the two terms in
$a_{4}^{(q)}-2r_{K}a_{6}^{(q)}$, the penguin emission
contributions of $B\rightarrow \rho^- K^+$ decay are no longer
leading. Instead, tree diagram may be the main effects. It is known that $%
a_{2}(\mu )$ is a flat function in $\mu $. However, $a_{4}^{(q)}(\mu )$ and $%
a_{6}^{(q)}(\mu )$ have an enormous enhancement at $\mu <M_{B}/2$ \cite{KLS1}%
. If $F_{a6}^{\rho K}$ is only few times smaller than
$F_{e}^{B\rho }$, with the penguin enhancement from
$a_{6}^{(q)}(\mu )$, the annihilation contributions could possibly
become the essential parts in $B\rightarrow \rho ^{-}K^{+}$ decay.
Recently, the estimation of annihilation topologies has been
proposed by MPQCD \cite{KLS1} and QCDF \cite{BBNS}. Although the
basic idea is originated from the Lepage and Brodsky (LB)
formalism \cite{LB}, in which a transition amplitude can be
factorized into the convolution of nonperturbative parts,
described by hadron wave functions, and a hard amplitude of
valence quarks, dictated by perturbative hard gluon exchanges,
both approaches have different description on factorization scale.
%%However, due to the different concept on factorization scale, consequently
%the typical scale of MPQCD is around $1.7$ GeV while that of QCDF is $%
%M_{B}/2 $. In this paper, we will concentrate on the MPQCD
%approach \cite
%{MPQCD}. As for the more detailed discussions on QCDF, one can refer to Refs.%
%\cite{BBNS,CY}.

It has been known that the original LB formalism suffers singularities from
the end-point region with a momentum fraction $x\to 0$.
In the QCDF approach, it is claimed that heavy $B$ meson decays at
the fast recoil region are still dominated by soft gluons, which
are uncalculable perturbatively. As a result, the decay amplitudes
for $B$ meson decays are written as $A\sim
C(t_0)f_{M_1}F^{BM_{2}}$ with $t_{0}\approx M_{B} \sim M_{B}/2$,
where $C(t_{0})$, $f_{M_1}$ and $F^{BM_{2}}$ are the relevant
wilson coefficient, decay constant of $M_{1}$ meson and
$B\rightarrow M_{2}$ form factor respectively. However, following
the concept of PQCD, if the spectator quark of inside $B$ meson,
carrying the momentum of order of $\bar{\Lambda}$ with
$\bar{\Lambda}=M_{B}-M_{b}$ and $M_{b}$ being $b$ quark mass,
wants to catch up with the outgoing quark, which is the daughter
of $b$ quark decay, to form a hadron, it should need to obtain a
large energy from $b$ quark or the daughter of $b$ quark. That is,
in contrast to the conclusion of QCF approach, hard gluons
actually play an essential role in $B$ meson decays. Therefore,
the relevant decay amplitude should be calculable perturbatively
and the order of magnitude of typical scale is similar to the
momentum of hard gluon approximately, denoted by
$\sqrt{\bar{\Lambda}M_{B}}$ \cite{KLS2,Chen}.

Now, how to deal with the problem of singularities is the main
part of PQCD.
In order to handle these singularities, the strategy of including
$k_{T}$, the transverse momentum of the valence quark, and
threshold resummation has been proposed \cite{MPQCD}. It has been
shown that the singularities do not exist in a self-consistent
MPQCD analysis \cite{MPQCD}. Additionally, MPQCD gives many
interesting results: for example, the $B\rightarrow P(V)$ form
factors are dominated by perturbative dynamics with $\alpha
_{s}/\pi <0.2$, the formalism involves
less theoretical uncertain parameters that arise from the shape parameter $%
\omega _{B}$ in the $B$ meson wave function, CSB parameter $m^0_{K}$ and the
parametrization of Sudakov factor from threshold resummation, the penguin
contribution is enhanced at a lower typical scale and annihilation
topologies contribute large absorptive parts. Especially, according to our
power counting rules, the ratios of the transition form factor ($F^{B\rho }$%
) to the annihilation contributions and to the nonfactorizable
contributions are found to be $=1:r_{K}$: $\bar{\Lambda}/M_{B}$
\cite{CKL}. For $M_{B}\sim 5.0$
GeV, the magnitude of the annihilation amplitude is only less than that of $%
F^{B\rho }$ by a factor of 3. In the literature, the applications
of the MPQCD to the processes of $B\rightarrow PP$, such as
$B\rightarrow K\pi $ \cite{KLS1}, $ B\rightarrow \pi \pi $
\cite{pipi}, $B\rightarrow KK$ \cite{CL}, $B\rightarrow
K\eta^{(\prime)}$ \cite{KS} and $B_{s}\rightarrow KK$ \cite{Chen},
 that of $B\rightarrow VP$ such as $B\rightarrow \phi
\pi $ \cite{Melic} and $B\rightarrow \rho (\omega) \pi $
\cite{LY}, as well as that of semileptonic decays such as $B\to
K^{(*)} \ell^+ \ell^{-}$,
 have been studied and found that they are consistent with the experimental data or
limits.

Hence, the $B\rightarrow \rho $ and time-like form factors, defined as
\begin{eqnarray*}
<\rho ^{-}(p_{\rho },\epsilon _{\rho })|\bar{b}\not{q}\gamma _{5}u|B(p_{B})>
&=&2m_{\rho }\epsilon _{\rho }^{*}\cdot qF_{e}^{B\rho }, \\
\left\langle \rho ^{-}(p_{\rho })K^{+}(p_{K})\right| \bar{u}\gamma
_{5}s\left| {\bf 0}\right\rangle &=&2m_{\rho }\epsilon _{\rho }^{*}\cdot
qF_{a}^{\rho K}
\end{eqnarray*}
with $q=p_{B}-p_{\rho }$, can be explicitly written as
\begin{eqnarray}
F_{e}^{B\rho } &=&8\pi
C_{F}M_{B}^{2}\int_{0}^{1}dx_{1}dx_{2}\int_{0}^{\infty
}b_{1}db_{1}b_{2}db_{2}  \nonumber \\ &&\phi _{B}\left(
x_{1},b_{1}\right) \left\{ \left[ \left( 1+x_{2}\right) \phi
_{\rho }\left( x_{2}\right) +r_{\rho }\left( 1-2x_{2}\right)
\right. \right.  \nonumber \\ &&\left. \left( \phi _{\rho
}^{t}\left( x_{2}\right) +\phi _{\rho }^{s}\left( x_{2}\right)
\right) \right] E_{e}\left( t_{e}^{\left( 1\right) }\right)
h_{e}\left( x_{1},x_{2},b_{1},b_{2}\right)  \nonumber \\ &&\left.
+\left[ 2r_{\rho }\phi _{\rho }^{s}\left( x_{2}\right) \right]
E_{e}\left( t_{e}^{\left( 2\right) }\right) h_{e}\left(
x_{2},x_{1},b_{2},b_{1}\right) \right\}  \label{fe4} \\
F_{a6}^{\rho K} &=&-8\pi
C_{F}M_{B}^{2}\int_{0}^{1}dx_{2}dx_{3}\int_{0}^{\infty
}b_{2}db_{2}b_{3}db_{3}  \nonumber \\ &&\left\{ \left[
x_{3}r_{K}\phi _{\rho }\left( x_{2}\right) \left( \phi
_{K}^{p}\left( x_{3}\right) -\phi _{K}^{\sigma }\left(
x_{3}\right) \right) \right. \right.  \nonumber \\ &&\left.
+2r_{\rho }\phi _{\rho }^{s}\left( x_{2}\right) \phi _{K}\left(
x_{3}\right) \right] E_{a}\left( t_{a}^{\left( 1\right) }\right)
h_{a}\left( x_{2},x_{3},b_{2},b_{3}\right)  \nonumber \\ &&+\left[
-r_{\rho }x_{2}\phi _{K}\left( x_{3}\right) \left( \phi _{\rho
}^{t}\left( x_{2}\right) -\phi _{\rho }^{s}\left( x_{2}\right)
\right) \right.  \nonumber \\ &&\left.\left. +2r_{K}\phi _{\rho
}\left( x_{2}\right) \phi _{K}^{p}\left( x_{3}\right) \right]
E_{a}\left( t_{a}^{\left( 2\right) }\right) h_{a}\left(
x_{3},x_{2},b_{3},b_{2}\right) \right\}  \label{fa6}
\end{eqnarray}
where $r_{\rho }=m_{\rho }/M_{B}$, $m_{\rho }$ is the $\rho $ meson mass, $%
C_{F}=4/3$ is the color factor, $x_{i}$ $(i=1,2,3)$ are the
momentum fractions carried by the spectator quarks inside the $B$,
$\rho $ and $K$ mesons, and the variables $b_{i}$ are conjugate to
parton transverse momenta $k_{iT}$. $\phi _{P(V)}(x)$ denote the
twist-2 P(V) meson wave functions, and $\phi _{P(V)}^{p(t)}(x)$
and $\phi _{P(V)}^{\sigma (s)}(x)$ correspond to the pseudoscalar
(tensor) and pseudotensor (scalar) twist-3 wave functions,
respectively \cite{CKL,KLS2}. The wave functions are pure
nonperturbative effects and universal. In our numerical
calculations, we will adopt the results of Refs. \cite{BBKT,Ball}
derived from QCD sum rules. The explicit expressions of hard
functions $h_{e(a)}$ are shown as
\begin{eqnarray}
h_{e}(x_{1},x_{2},b_{1},b_{2}) &=&K_{0}\left( \sqrt{x_{1}x_{2}}
M_{B}b_{1}\right)S_t(x_2) \nonumber \\ &&\times \left[ \theta
(b_{1}-b_{2})K_{0}\left( \sqrt{x_{2}} M_{B}b_{1}\right)
I_{0}\left( \sqrt{x_{2}}M_{B}b_{2}\right) \right. \nonumber \\
&&\left. +\theta (b_{2}-b_{1})K_{0}\left(
\sqrt{x_{2}}M_{B}b_{2}\right) I_{0}\left(
\sqrt{x_{2}}M_{B}b_{1}\right) \right] \;, \label{he}
\\ h_{a}(x_{2},x_{3},b_{2},b_{3}) &=&\left( \frac{i\pi
}{2}\right)^{2} H_{0}^{(1)}\left(
\sqrt{x_{2}x_{3}}M_{B}b_{2}\right)S_t(x_3) \nonumber \\ &&\times
\left[ \theta (b_{2}-b_{3})H_{0}^{(1)}\left( \sqrt{x_{3}}
M_{B}b_{2}\right) J_{0}\left( \sqrt{x_{3}}M_{B}b_{3}\right)
\right. \nonumber \\ &&\left. +\theta
(b_{3}-b_{2})H_{0}^{(1)}\left( \sqrt{x_{3}} M_{B}b_{3}\right)
J_{0}\left( \sqrt{x_{3}}M_{B}b_{2}\right) \right] \;, \label{ha}
\end{eqnarray}
where $S_{t}(x)$ is the evolution function from threshold
resummation parametrized by $S_{t}\approx N_{t}\left[
x(1-x)\right] ^{c}$ \cite{MPQCD}, and $K_0, I_0, H_0$ and $J_0$
are the Bessel functions. The evolution factors are given by
\begin{equation}
E_{e(a)}(t)=\alpha _{s}(t)\exp [-S_{B(K
)}(t,x_{1(3)})-S_{\rho}(t,x_{2})]\,, \label{ea}
\end{equation}
which contain the Sudakov factor from $k_T$ resummation \cite{CS,BS}.

Adopting $f_{B}=0.19$, $f_{\rho }=0.20$, $f_{K}=0.16$ GeV and with the
allowed values of parameters, we obtain $F_{e}^{B\rho }=0.37_{-0.02}^{+0.03}$%
. Via the same procedure, we get $F_{e}^{BK}=0.35_{-0.02}^{+0.04}$ \cite{CKL}
and $F_{e}^{B\pi }=0.3_{-0.02}^{+0.03}$ \cite{CKL,KLS2}. All of these form
factors are consistent with those from light-cone QCD sum rules (LCSR) \cite
{Ali} and quark model (QM) \cite{MS}. Similarly, the time-like form factor
is given by $F_{a6}^{\rho K}=\left( -0.46+7.4i\right) \times 10^{-2}$, in
which the large absorptive part arises from the on-shell condition of
internal light quarks. As expected, the magnitude of $F_{e}^{B\rho }$ is
only larger than that of $F_{a6}^{\rho K}$ by a factor of $5$ in the MPQCD
approach. In terms of the above results, from Eq. (\ref{rhok}), if choosing $%
\gamma\simeq 90^0$, the contributions of tree and penguin
annihilation are comparable but are opposite in sign, such that
the BR of $B\rightarrow \rho ^{-}K^{+}$ is small. However, via
$V_{u}$ instead of $V_{u}^{*}$, we find immediately that the CP
conjugate mode $B\rightarrow \rho ^{+}K^{-}$ is enhanced. It is
interesting to question how large the BR can be pushed up by such
enhancement. For estimation, we assume $F_{a6}^{\rho K}\approx
F_{a6}^{\phi K}$ and use the above obtained values, from Eqs.
(\ref{phik}) and (\ref{rhok}), the relation $\left|
A(B^{+}\rightarrow \phi K^{+})\right| \sim 1.17$ $\left|
A(B\rightarrow \rho ^{+}K^{-})\right| $ for $\gamma
\simeq 90^0$ and for $\mu \approx 1.7$ GeV is obtained. With $%
BR(B^{+}\rightarrow \phi K^{+})\approx 10\times 10^{-6}$, we get $%
BR(B\rightarrow \rho ^{+}K^{-})\sim 7\times 10^{-6}$. Furthermore, due to
the destruction between the tree and penguin amplitudes in $B\rightarrow
\rho ^{-}K^{+}$ decay, one also expects the CPA in $B\rightarrow \rho ^{\mp
}K^{\pm }$ is large.

We have studied the properties of $B\rightarrow \rho ^{\mp }K^{\pm }$ and
found that by the constructive interference between the tree and penguin
annihilation topologies, the BR and CPA estimated in the MPQCD are not as
small as those expected in Ref. \cite{AC}. As mentioned before, basically
all the modes in $B\rightarrow \rho (\omega )K$ have the similar properties.
In addition, for those decays involving $\rho ^{0}$ and $\omega $ mesons
that are composed by $(\bar{u}u-\bar{d}d)/\sqrt{2}$ and $(\bar{u}u+\bar{d}d)/%
\sqrt{2}$, respectively, the new terms $%
(a_{35}^{(u)}-a_{35}^{(d)})F_{e}^{BK} $ and $%
(a_{35}^{(u)}+a_{35}^{(d)})F_{e}^{BK}$ with $%
a_{35}^{(q)}=a_{3}^{(q)}+a_{5}^{(q)}$ will be induced. Because the values of
$a_{3}^{(q)}$ and $a_{5}^{(q)}$ are much smaller than those of $%
a_{4,6}^{(q)} $, the properties discussed before will not change.
Furthermore, we can separate these decays into two classes. Class
I consists of $B\rightarrow \rho ^{\mp }K^{\pm }$, $B^{\mp
}\rightarrow \rho ^{0}K^{\mp }$ and $B^{\mp }\rightarrow \omega
K^{\mp }$, while $B\rightarrow \rho ^{0}K^{0}$, $B\rightarrow
\omega K^{0}$ and $B^{\mp }\rightarrow \rho ^{\mp }K^{0}$ are in
class II. We find that the tree contributions in class I are
related to $a_{2}F_{e}^{B\rho } $, but in class II they are associated with $%
a_{1}F_{e}^{BK}$ or $a_{2}F_{a4}^{\rho K}$. As known, $a_{1}$ is one order
of magnitude less than $a_{2}$ and $F_{a4}^{\rho K}$ is much smaller than $%
F_{a6}^{\rho K}$, so that the tree effects associated with $%
V_{u}^{*}a_{1}F_{e}^{BK}$ and $V_{u}^{*}a_{2}F_{a4}^{\rho K}$ are quite
small. Thus, we expect that the large CPAs and BRs exist only in the modes
of class I. We emphasize that the dynamical variables $a_{i}^{(q)}(\mu )$ in
the MPQCD approach have to be as a evolution factor convoluted with $%
F_{e[a]}^{P(V)P}$. The full formulas can be easily obtained by inserting the
relevant $a_{i}^{(q)}(\mu )$ to Eq. (\ref{ea}) \cite{CKL} and setting $\mu
=t $. In our following numerical calculations, we will adopt such formulas.

\begin{center}
\begin{figure}[tbp]
\centerline{\psfig{figure=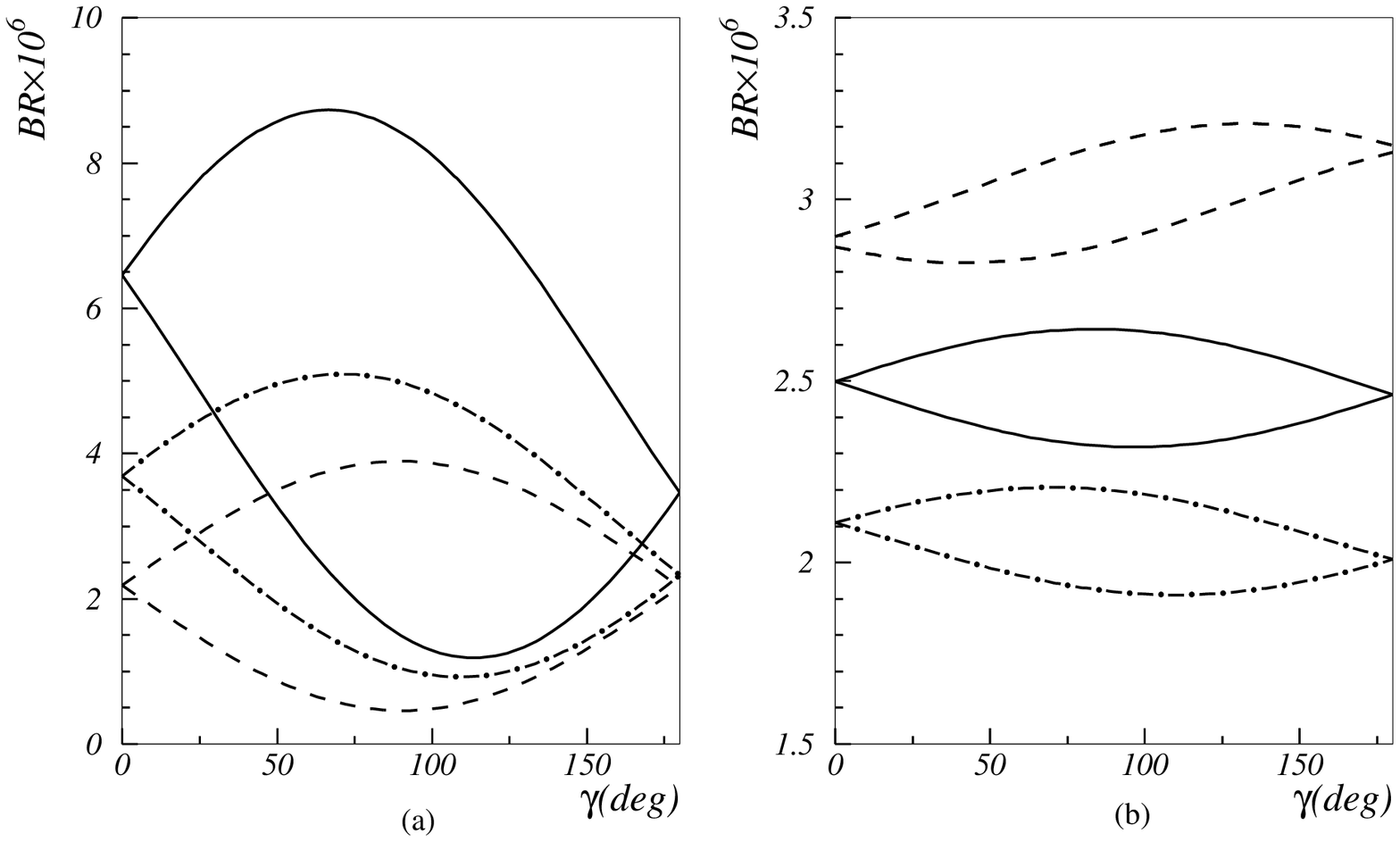,height=3.in }} \caption{BRs of
the $B\rightarrow\rho(\omega) K$ decays as a function of the
angle $\gamma$. The solid, dashed and dash-dotted lines are for (a) $%
B_d\rightarrow\rho^{\pm}K^{\mp}$, $B^{\mp}\rightarrow\rho^{0}K^{\mp}$, and $%
B^{\mp}\rightarrow\omega K^{\mp}$, and for (b) $B_d\rightarrow\rho^{0}K^{0}$%
, $B^{\pm}\rightarrow\rho^{\pm}K^{0}$, and $B_d\rightarrow\omega K^0$,
respectively. }
\label{figurebr}
\end{figure}
\end{center}

With the data of $V_{us}\simeq \lambda $, $V_{ts}\approx -A\lambda ^{2}$, $%
V_{ub}\approx A\lambda ^{3}R_{b}e^{-i\gamma}$, $A\approx 0.81$, $\lambda $ $%
\approx 0.22$, $R_{b}\approx 0.36$, $N_{t}\approx 1.77$, $c \approx 0.3$, $%
m_{K}^{0}=1.7$ and $\omega _{B}=0.4$ GeV as the input values and
considering all factorizable and nonfactorizable effects, the
results of BRs with CP average and CPAs are shown in Table
\ref{tablebr}. Note that the CPAs in class I are large. That means
the BR of one of decay modes much larger than that of its CP
conjugate mode. For illustration, they are given as ($
B\rightarrow \rho ^{+ }K^{- }$, $B^{- }\rightarrow \rho ^{0}K^{-
}$, $B^{-}\rightarrow \omega K^{-}$)$=$($8.72$, $3.81$,
$5.10$)$\times
10^{-6}$ for $\gamma \simeq 72^{0}$. We also plot the BRs as a function of $%
\gamma $ in Figure \ref{figurebr}. It is worth of pointing out
that except tree effects, the $B^{\mp }\rightarrow \rho ^{\mp
}K^{0}$ and $B\rightarrow \rho ^{-}K^{+}$ decays have the similar
contributions from the penguin topologies. According to our
estimation, the effect of penguin emission associated with $\left(
a_{4}^{(d)}-2r_{K}a_{6}^{(d)}\right) F_{e}^{B\rho }$ on the BR is
only around $19\%$. Therefore, the value of $BR$($B^{\mp
}\rightarrow \rho ^{\mp }K^{0}$) in Table \ref{tablebr} almost
indicates the effects of annihilation contributions. In MPQCD, the
uncertainties of hadronic effects can be  from the power factor
$c$ for the parametrization of threshold resummation, $\omega_{B}$
and $m^0_{K}$. With the allowed regions
\cite{KLS1}, we find that the uncertainties of power factor $c$ and $%
\omega_{B}$ are below $10\%$ and the error of $m^0_{K}$ is $20\%$.
The most challenge contributions are from higher order
corrections. In order to estimate their size, we fix all free
parameters to specific values, shown in the beginning of this
paragraph, and then change the chosen condition of $t$ scale by
 a deviation of $30\%$. We find that the influence on BRs is also
around $30\%$.

%In the following,
We now give a brief discussion on the nonfactorizable effects. It is known
that besides %$a'^{(q)}_{4,6}$
contributions from $a^{\prime(q)}_{4,6}$ to the nonfactorizable
diagrams of the hard gluon exchanges between the valence quarks in
$\rho$ and $B$ mesons, those from $a^{\prime
(u)}_{35}-a^{\prime(d)}_{35}$ and $a^{\prime
(u)}_{35}+a^{\prime(d)}_{35}$ for $\rho^0$ and $\omega$ modes will
also be introduced. According to Eq. (\ref{ew}), the former is
only associated with the small WCs of $C_{7,10}(\mu)$ but the
latter is related to $C_{4,6}$. Thus, we expect that the influence
of nonfactorizable effects on $\omega K$ modes is extraordinary.
From our estimations, by neglecting all nonfactorizable
contributions, we find that the Brs of $B\rightarrow \omega
K^{\mp} (K^{0})$ are reduced about $50\%$, while the influence on
$Br( \rho K^{\mp})$ and $Br( \rho K^{0})$ are around $35\%$ and
$18\%$, respectively.

Finally, we give a remark on the QCDF approach. According to the analysis of
Ref. \cite{CY}, the BR of $B^{\pm}\rightarrow \phi K^{\pm}$ can reach $%
7.3\times 10^{-6}$, if the included annihilation effects are almost real. If
taking this value as the common result of MPQCD and QCDF, with the similar
procedure, the values of the CP-averaged BR for $B\rightarrow \rho (\omega)
K $ in QCDF can be estimated as ($B\rightarrow \rho^{\mp} K^{\pm}$, $%
B^{\mp}\rightarrow \rho^{0} K^{\mp}$, $B^{\mp}\rightarrow \omega K^{\mp}$)$%
\sim$ (2.51, 0.62, 1.28) $\times 10^{-6}$ and ($B^{\mp}\rightarrow
\rho^{\mp} K^{0}$, $B\rightarrow \rho^{0} K^{0}$, $B\rightarrow
\omega K^{0}$)$\sim$ (1.63, 1.99, 0.64) $\times 10^{-6}$ for
$\gamma\simeq 72^0$. The results in MPQCD are $35\%$ smaller than
those in Table \ref{tablebr}. Though some results are similar in
both approaches, the CPAs from QCDF for class I are found to be
only at the few percent level. If the considered annihilation
effects are almost imaginary, the CPAs from QCDF for class I can
be as large as those from MPQCD. However, the BRs from QCDF become
much smaller than those from MPQCD. Altogether, even if MPQCD and
QCDF have the common prediction for $B\rightarrow \phi K$, it is
clear that $B\rightarrow \rho (\omega) K$ can distinguish the
different QCD approaches.

In summary, we have performed the analysis of the $B\rightarrow \rho (\omega
)K$ decays. Although the unique dynamical penguin enhancement in MPQCD can
be verified by the measurements of $B\rightarrow \phi K$, which distinguish
MPQCD from QCDF, $B\rightarrow \rho (\omega )K$ can display not only such an
enhancement but also the importance of annihilation topologies, especially
their absorptive parts. We also show that nonfactorizable effects play an
important role in $B\rightarrow \omega K$ decays. Moreover, due to the
cancellation in $a_{4}^{(q)}-2r_{K}a_{6}^{(q)}$ for penguin emission
topologies, any sizable new physics, with or without new weak CP violating
phases contributing to the penguin operators ${\cal O}_{3}\sim {\cal O}_{10}$%
, will have a remarkable influence on the BRs and CPAs of the modes in class
I and II. $B\rightarrow \rho (\omega )K$ are also sensitive to physics
beyond standard model. Recently, CLEO has reported results of $B\rightarrow
\rho (\omega )K$, such as $BR$($B\rightarrow \rho ^{\mp }K^{\pm } $)$=$($%
16.0_{-6.4}^{+7.6}\pm 2.8$)$\times 10^{-6}$ and $BR$($B^{\mp
}\rightarrow
\omega K^{\mp }$)$=$($3.2_{-1.9}^{+2.4}\pm 0.8$)$\times 10^{-6}$ \cite{CLEO1}%
. It is obvious that the central value of $BR$($B^{\pm }\rightarrow \omega
K^{\pm }$) is the same as our prediction. The further confirmation will be
made, when the data from the B factories are announced in the near future.
\newline

\noindent {\bf Acknowledgments:}

The author would like to thank H.Y. Cheng, C.Q. Geng, X.G. He,
W.S. Hou, Y.Y. Keum, H.N. Li, and K.C. Yang for their useful
discussions. This work was supported in part by the National
Science Council of the Republic of China under Grant No.
NSC-89-2112-M-006-033 and by the National Center for Theoretical
Science.

\newpage

\begin{table}[htp]
\caption{ BRs (in units of $10^{-6}$) with CP average for class I
and II respectively and CP asymmetries (\%) in the $B\rightarrow
\rho(\omega)K$ decays for $\gamma\simeq72^0$. Nonfactorizable
effects are included.} \label{tablebr}
\begin{center}
\begin{tabular}{lrll}
\hline &  & Class I &  \\ \hline Mode &
\multicolumn{1}{l}{$B\rightarrow \rho ^{\pm}K^{\mp}$} &
$B^{\mp}\rightarrow
\rho ^{0}K^{\mp}$ & $B^{\mp}\rightarrow \omega K^{\mp}$ \\
BR & \multicolumn{1}{c}{$5.42$} & \multicolumn{1}{c}{$2.18$} &
\multicolumn{1}{c}{$3.22$} \\
CPA & \multicolumn{1}{c}{$-60.76$} & \multicolumn{1}{c}{$-74.95$} &
\multicolumn{1}{c}{$-58.21$} \\ \hline
& \multicolumn{1}{l}{} & Class II &  \\ \hline
Mode & \multicolumn{1}{l}{$B^{\pm }\rightarrow \rho^{\pm }K^{0}$} & $%
B\rightarrow \rho ^{0}K^{0}$ & $B\rightarrow \omega K^{0}$ \\
BR & \multicolumn{1}{c}{$2.96$} & \multicolumn{1}{c}{$2.49$} &
\multicolumn{1}{c}{$2.07$} \\
CPA & \multicolumn{1}{c}{$-3.74$} & \multicolumn{1}{c}{$-6.18$} &
\multicolumn{1}{c}{$6.39$} \\ \hline
\end{tabular}
\end{center}
\end{table}

\end{document}